\documentclass[final]{siamart220329}

\usepackage{lipsum}
\usepackage{amsfonts}
\usepackage{graphicx}
\usepackage{epstopdf}
\usepackage{algorithmic}
\usepackage{derivative}
\usepackage{caption}
\usepackage{subcaption}

\usepackage{physics}
\usepackage{tikz}
\usepackage{multirow}

\usepackage{acronym}
\usetikzlibrary{shapes, arrows.meta, positioning}
\usepackage{float}
\usepackage{placeins}
\ifpdf%
  \DeclareGraphicsExtensions{.eps,.pdf,.png,.jpg}
\else
  \DeclareGraphicsExtensions{.eps}
\fi
\usepackage{amsopn}

\usepackage{booktabs}

\newcommand{\TheTitle}{%
  Evolving Algebraic Multigrid Methods Using Grammar-Guided Genetic Programming
}

\newcommand{\TheShortTitle}{%
  Evolving AMG Methods Using Grammar-Guided Genetic Programming 
}

\newcommand{\TheName}{%
  Dinesh Parthasarathy
}

\newcommand{\TheAddress}{%
  University of Erlangen-Nuremberg,
  (\email{dinesh.parthasarathy@fau.de}).
}

\newcommand{\Theconference}{%
 Presented at the Copper Mountain Conference on Iterative Methods, April 2024. 
}

\newcommand{\TheCollaborators}{%
  Wayne Bradford Mitchell,
  Harald Köstler,
}

\author{\TheName\thanks{\TheAddress}}
\title{{\TheTitle}\thanks{\Theconference}}
\headers{\TheShortTitle}{\TheName}
\ifpdf%
\hypersetup{%
  pdftitle={\TheTitle},
  pdfauthor={\TheName}
}
\fi

\acrodef{FAS}{full approximation scheme}
\acrodef{DSL}{domain specific language}
\acrodef{GP}{Genetic programming}
\acrodef{G3P}{Grammar-guided genetic programming}
\acrodef{CFG}{context-free grammar}
\acrodef{PDE}{partial differential equation}
\acrodef{EA}{Evolutionary Algorithm}
\acrodef{AI}{artificial intelligence}
\acrodef{AMG}{algebraic multigrid}
\acrodef{GMG}{geometric multigrid}
\acrodef{CG}{conjugate gradient}
\acrodef{PCG}{preconditioned conjugate gradient}
\acrodef{AMG-PCG}{AMG preconditioned CG}
\acrodef{CGC}{coarse-grid correction}
\begin{document}

\maketitle

\begin{center}
In collaboration with:
  {\TheCollaborators}
\end{center}
\vspace{1cm}

\begin{abstract}
Multigrid methods despite being known to be asymptotically optimal algorithms, depend on the careful selection of their individual components for efficiency. Also, they are mostly restricted to standard cycle types like V-, F-, and W-cycles. We use grammar rules to generate arbitrary-shaped cycles, wherein the smoothers and their relaxation weights are chosen independently at each step within the cycle. We call this a flexible multigrid cycle. These flexible cycles are used in Algebraic Multigrid (AMG) methods with the help of grammar rules and optimized using genetic programming. The flexible AMG methods are implemented in the software library of hypre, and the programs are optimized separately for two cases: a standalone AMG solver for a 3D anisotropic problem and an AMG preconditioner with conjugate gradient for a multiphysics code. We observe that the optimized flexible cycles provide higher efficiency and better performance than the standard cycle types.
\end{abstract}


\section{Introduction}\label{sec:intro}
\begin{figure}
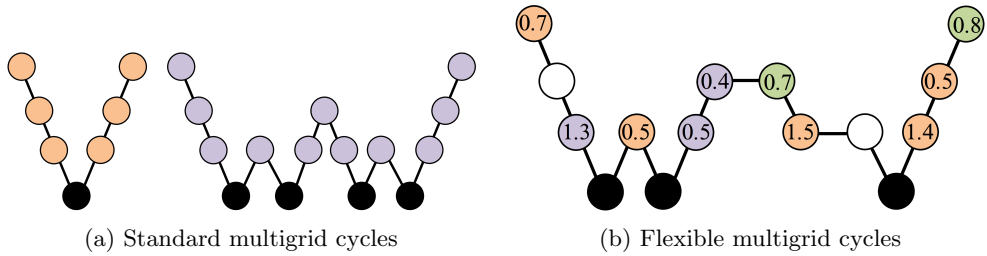

    \begin{subfigure}{0.48\textwidth}
        \centering
        \includegraphics[width=\linewidth]{images/standardMGcycles.pdf}
        \caption{Standard multigrid cycles}
        \label{fig:stdmgcycles}
    \end{subfigure}\hfill
    \begin{subfigure}{0.48\textwidth}
        \centering
        \includegraphics[width=\linewidth]{images/FlexMG.pdf}
        \caption{Flexible multigrid cycles}
        \label{fig:flexmgcycles}
    \end{subfigure}
    \caption{Visual representation of the cycling structure in a multigrid method. The color of the node denotes the smoother type with its relaxation weights indicated inside. }
    \label{fig:cyclestruct}
\end{figure}

Multigrid methods are a class of highly efficient algorithms for solving large systems of discretized partial differential equations. The selection of algorithmic components in a multigrid method plays an important role in determining its efficiency. The task of designing an efficient multigrid method is a non-trivial one. Leveraging the recent advances in \ac{AI}, efforts have been made to find optimal multigrid components, such as smoothers \cite{Huang2023-jb}, intergrid operators \cite{pmlr-v97-greenfeld19a,10.5555/3524938.3525540, Katrutsa2020, katrutsa2017deep}, and coarsening schemes \cite{Taghibakhshi2021OptimizationBasedAM}.  We take a complementary approach and construct efficient multigrid cycles from a set of available multigrid components. Traditional multigrid methods employ recursive cycle types such as V-, W-, F, or the more recent $\kappa$-cycles \cite{Avnat2023-nr}. However, we use a so-called \textit{flexible multigrid cycle}, which can arbitrarily cycle up or down in a non-recursive manner. Also, the choice of smoothers and their relaxation weights can be made individually for each step in the cycle (Fig \ref{fig:flexmgcycles}). These flexible cycles are challenging to hand-tune due to their extensive search space, but they can be formulated as a program optimization task. This way, we aim to exploit the inherent flexibility in the cycling structure, with the intent of discovering methods more efficient than the standard cycling types. Schmitt et al. use a similar approach to construct efficient grammar-guided \ac{GMG} methods \cite{Schmitt2021, Schmitt2022}. This involves the automatic generation of \ac{GMG} programs evolved using genetic programming. We adapt this methodology to construct efficient \ac{AMG} methods. This approach is used to optimize an \ac{AMG} method as a solver for a 3D anisotropic problem and then as a preconditioner with \ac{PCG} for a multiphysics simulation. The constructed methods are evaluated and compared to optimized reference methods with standard \ac{AMG} cycling.

\FloatBarrier

\section{Background}\label{sec:back}
\ac{GP} is a research area in \ac{AI} that deals with the evolution of computer code. It is one of the branches of \acp{EA}, following Darwin's principle of natural selection. The general approach of \ac{GP} is to apply a population of programs to a given problem, and compare their performance (fitness) relative to each other. Operators inspired by genetics (crossover, mutation) are applied to selected programs from the population (parents) such that in time better programs (offsprings) emerge by evolution. This principle is applied iteratively for multiple generations until a desirable population of programs is obtained \cite{Banzhaf2001, Orlov2009}. 
In problems where the programs are required to conform to a specific structure (for eg. multigrid methods), grammar rules can be used to constrain the evolutionary process such that only valid programs are produced. \ac{G3P} is an extension of traditional \ac{GP} systems that use \acp{CFG} to impose such constraints on the initial population and subsequent genetic operations. Exploiting domain knowledge using \acp{CFG} helps in eliminating invalid programs and speeds up the optimization process \cite{Whigham1995GrammaticallybasedGP, Manrique2009}.

\section{Method}\label{sec:method}
Consider a system of linear equations $Au=f$, with a coarser level system $A_cu_c=f_c$, for the residual equations. Every step during the solve phase of an \ac{AMG} method can be written in the form
$v^{i+1} \leftarrow v^i + \omega B(f-Av^i)$, where $v^i$ is an approximate solution at iteration $i$, $B$ is an operator and $\omega \in \mathbb{R}$ is a relaxation factor for a smoothing step or a scaling factor for a \ac{CGC} step. For instance, a Jacobi smoothing step can be defined as $v^{i+1} \leftarrow v^i + \omega D^{-1}(f-Av^i)$, where $D=diag(A)$. Similarly, the \ac{CGC} step can be expressed as $v^{i+2} \leftarrow v^{i+1} + \omega P{A_c}^{-1}P^T(f-Av^{i+1})$. Here, $P$ is the interpolation operator that maps the coarse system to the original system of equations and $P^T$ is the transpose that maps from the original system to the coarser level. Substituting $v^{i+1}$ from the previous expression here, yields a single expression for a two-grid \ac{AMG} method with Jacobi pre-smoothing. This approach is extensible to multiple levels, and substituting subexpressions recursively creates a single final expression that maps uniquely to an arbitrary flexible multigrid cycle (Fig. \ref{fig:flexmgcycles}). Schmitt et al. designed a \ac{CFG} for multigrid methods defining a list of rules to generate such expressions automatically \cite{Schmitt2021}. We use this \ac{CFG} to generate different multigrid cycles for a given \ac{AMG} setup. This lets us automatically generate an initial population of \ac{AMG} programs and further discover optimal \ac{AMG} cycles using \ac{G3P}.

\section{Implementation}\label{sec:implementation}
\begin{figure}
    \centering
    \includegraphics[width=\textwidth]{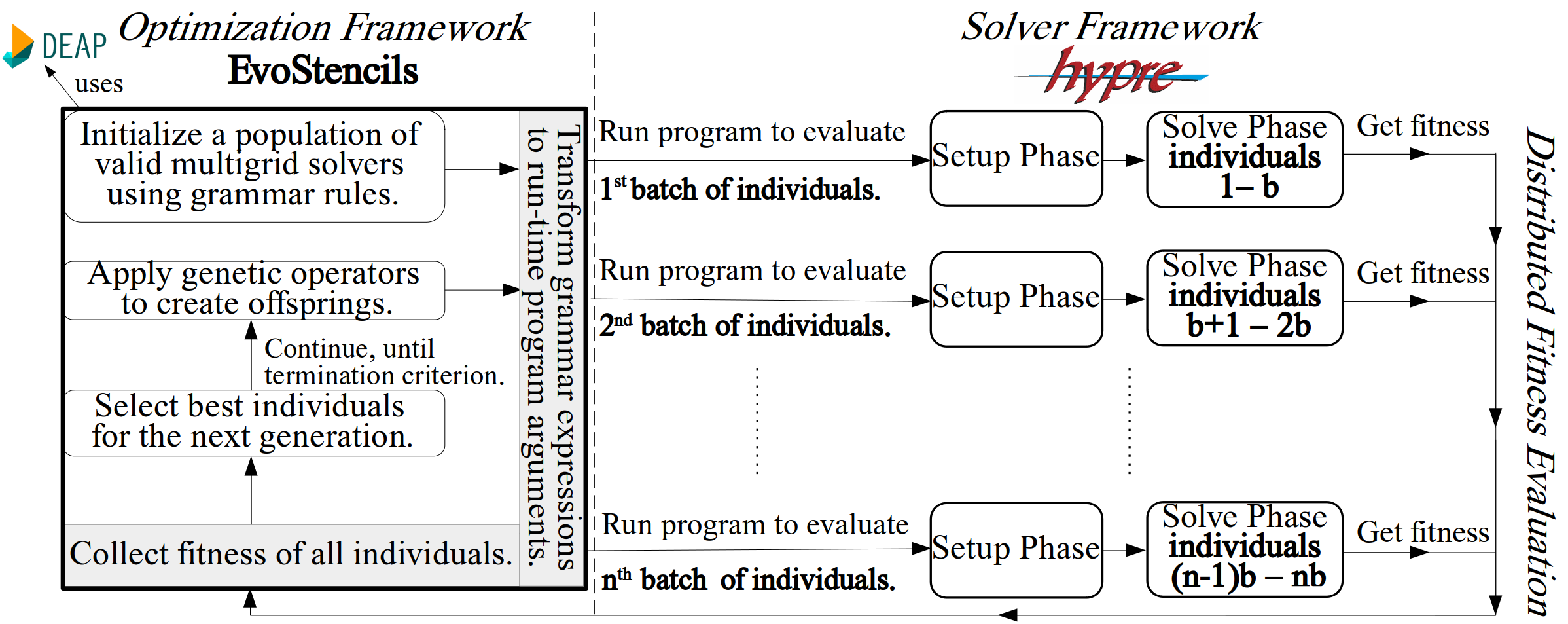}
    \caption{Software setup}
    \label{fig:softwareflow}
\end{figure}

The \textit{BoomerAMG}\footnote{https://hypre.readthedocs.io/en/latest/solvers-boomeramg.html} implementation from the software library \textit{hypre} is used to generate the AMG programs. Additional interfaces are added and implemented so that \ac{AMG} methods with flexible cycling can be defined and used within the \textit{hypre} framework\footnote{https://github.com/dinesh-parthasarathy/hypre}. The \ac{AMG} expression generated from the \ac{CFG} is transformed to corresponding input arguments for \textit{BoomerAMG}. Executing each program returns the \textit{solve time} and \textit{convergence factor} as the fitness measure for the optimization. The optimization is performed using the EvoStencils framework\footnote{https://github.com/jonas-schmitt/evostencils}. This uses the \ac{CFG} introduced earlier to generate high-level representations of multigrid methods, and the DEAP library\footnote{https://deap.readthedocs.io/en/master/} for applying evolutionary algorithms. We adapt this framework to include \ac{AMG} components from \textit{hypre} during expression generation. To eliminate redundant \ac{AMG} setup times during fitness evaluation, the program individuals are grouped into batches of size $b$, with one setup phase per batch, and then the solve phase for the individual solvers is executed in succession (Fig. \ref{fig:softwareflow}). The batches are distributed across multiple processes using MPI bindings\footnote{https://mpi4py.readthedocs.io}. 

\section{Experimental Setup}
We evaluate our approach to discover efficient \ac{AMG} methods when used as solvers and preconditioners. During the optimization, the \ac{AMG} setup phase is fixed. From a given choice of smoothers, a set of relaxation weights for smoothing, a set of scaling factors for \ac{CGC}, and flexibility in the cycling structure, we search for optimal \ac{AMG} methods (Table \ref{tab:amgcomp}a). We enable cycle flexibility for the top five levels and use a $V(1,1)$ \footnote{with Gauss-Seidel forward and backward solve on the up cycle and down cycle} cycle on the coarser levels. This allows the optimized methods to be used for different problem sizes. The methods are optimized for two objectives\textemdash\textit{solving time per iteration} and \textit{convergence factor}, and a 2D Pareto front is constructed. Eventhough we finally choose methods with the fastest total time to solution, using two objectives maintains diversity in the population and avoids premature convergence. Using separate objectives ensures that both cheap methods (per iteration), and methods with a high convergence rate (but probably expensive) are found in the optimization. The optimization is performed on the Ruby compute cluster at Lawrence Livermore National Laboratory\footnote{https://hpc.llnl.gov/hardware/compute-platforms/ruby}. Each node has two 28-core Intel Xeon CLX-8276L processors. The fitness evaluation for the program population is distributed across 32 nodes, spawning 64 MPI processes pinned to each processor (Table \ref{tab:g3pparam}b). The individual programs in each processor are parallelized across 28 cores using OpenMP. The following two cases are evaluated using this setup. 
\begin{table}
  \caption{Overview of the optimization setup}
  \begin{flushleft}
    \begin{minipage}{0.6\textwidth} 
      \begin{tabular}{r|p{0.5\textwidth}}
        \textit{Smoothers} & \begin{tabular}[t]{@{}l@{}}Gauss-Seidel forward,\\ Gauss-Seidel backward,\\ Jacobi\end{tabular}\\
        \textit{Relaxation weights} & (0.1, 0.15, 0.2, ..., 1.9)\\
        \textit{Scaling factors} & (0.1, 0.15, 0.2, ..., 1.9)\\
        \textit{Coarsening strategy} & HMIS algorithm \cite{DeSterck2007}\\
        \textit{Interpolation} & Extended+i  \cite{DeSterck2006} \\
        \textit{Restriction} & Interpolation transpose\\
        \textit{Coarse grid solver} & Gaussian elimination\\
        \label{tab:amgcomp}
      \end{tabular}
      \subcaption{\ac{AMG} components for the optimization}
    \end{minipage}%
    \begin{minipage}{0.4\textwidth} 
      \begin{tabular}{r|p{0.4\textwidth}}
        \begin{tabular}[t]{@{}l@{}}\textit{Evolutionary} \\ \textit{algorithm}\end{tabular}  & ($\mu + \lambda$) \\
        \textit{Generations}  & 100 \\
        \textit{$\mu$(population)}  & 256 \\
        \textit{$\lambda$(offsprings)}  & 256 \\
        \textit{Initial pop.}  & 2048 \\
        \textit{MPI processes}  & 64 \\
        \textit{Sorting alg.}  & NSGA-II \\
        \label{tab:g3pparam}
      \end{tabular}
      \subcaption{\ac{G3P} parameters}
    \end{minipage}
  \end{flushleft}
\end{table}

\subsection{AMG as a solver}\label{subsec:amgsolver}
\ac{AMG} method as a standalone solver is optimized for the following 3D anisotropic problem:
\begin{equation}
\begin{aligned}
-au_{xx}-bu_{yy}-cu_{zz} &= f \ \ \text{in} \ \Omega, \\
u &= 0 \ \ \text{on} \ \partial\Omega
\end{aligned}
\end{equation}
The system of equations is built using a standard 7-point laplacian stencil in a $100\times100\times100$ grid with anisotropy $a=0.001,b=1,c=1$ and $f=0$. The system is considered solved when the initial defect is reduced by a factor of $10^{-8}$.
\subsection{AMG as a preconditioner}\label{subsec:amgprecond}
Here, we find optimal \ac{AMG} methods as a preconditioner for \ac{PCG}. The \ac{PCG} method with \ac{AMG} preconditioner is used to solve an application in the Ares framework, a parallel multiphysics code developed and maintained at Lawrence Livermore National Laboratory. The \ac{AMG} method is optimized using the system matrix at a specific time step, and later evaluated on other time instances (with slightly different matrices than the one used for optimization). The performance is compared to a \ac{PCG} method with optimized standard \ac{AMG} cycling.
\section{Evaluation}\label{sec:eval}
The experiments for each of the two cases mentioned are conducted multiple times to ensure the robustness and consistent convergence of the optimization process towards an optimal set of solutions. Additionally, the newly discovered optimal \ac{AMG} methods are evaluated and compared to standard reference \ac{AMG} methods.
\subsection{AMG as a solver}
\begin{figure}
  \centering
  \includegraphics[width=\textwidth]{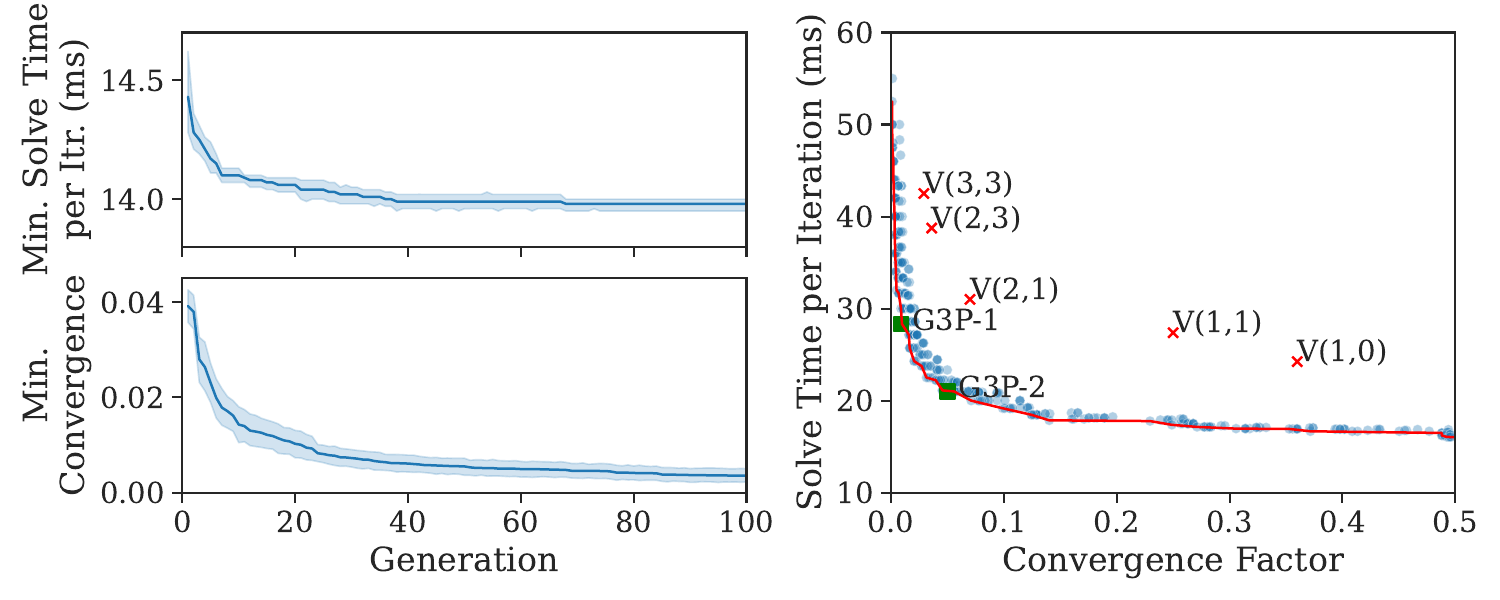}
  \caption{Mean and standard deviation of the minimum objective function values during the optimization (left) and combined Pareto distributions (right), over ten experiments.
}
  \label{fig:amgsolver_optimization}
\end{figure}

\begin{figure}
  \centering
  \includegraphics[width=\textwidth]{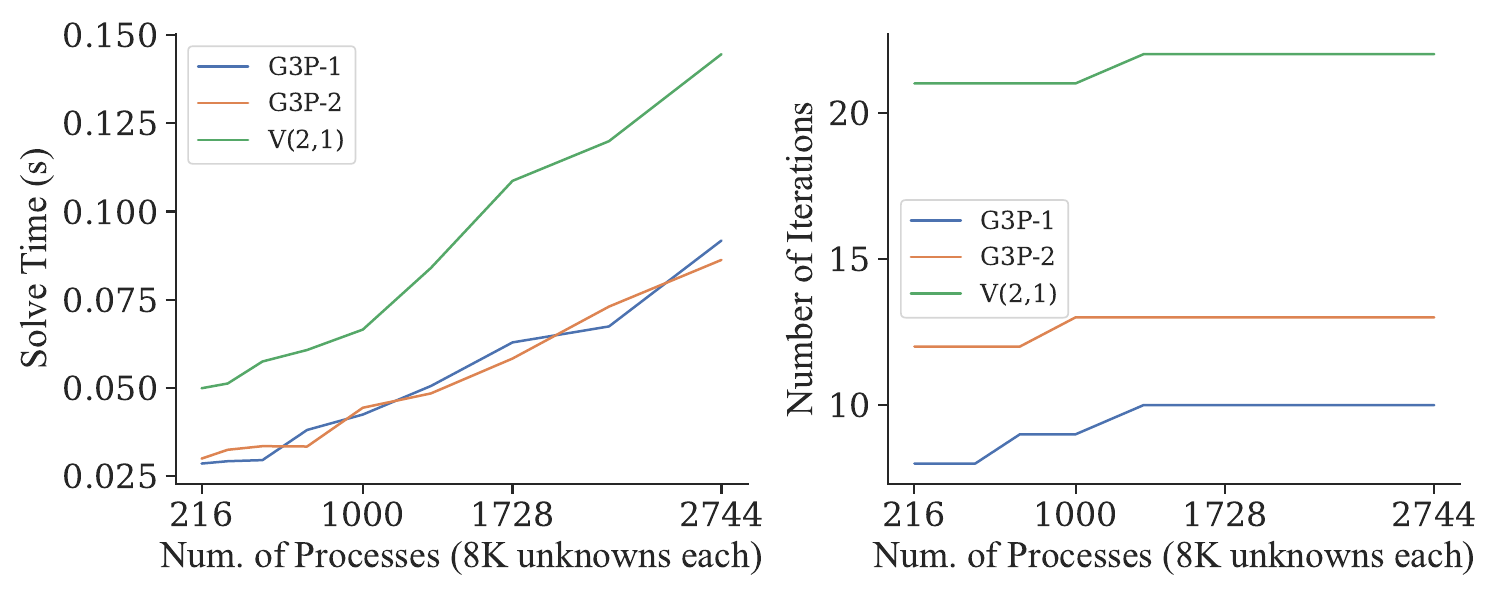}
  \caption{Weak scaling of a 3D anisotropic poisson problem.
}
  \label{fig:weakscaling_poisson}
\end{figure}

The optimization process is executed over 100 generations, and the evolution of objective values is monitored. It is observed that the minimum values of both objectives within the population decrease over time (Fig. \ref{fig:amgsolver_optimization}, left). This implies the discovery of \ac{AMG} methods that are not only computationally efficient but also converge rapidly. However, to minimize the total time required to reach a solution, a suitable \ac{AMG} method should reasonably exhibit both of these properties concurrently. Throughout the optimization process, Pareto optimal solutions from each generation are retained for further evolution. As a result, upon termination of the optimization, we obtain a diverse set of Pareto optimal solutions, each showcasing one or both of the desired traits. It is noteworthy that standard \ac{AMG} cycles are found to lie outside the Pareto front, indicating their suboptimal nature (Fig. \ref{fig:amgsolver_optimization}, right). \\ From the pool of Pareto optimal solutions, we select two well-performing solvers: i. G3P-1, a fast-converging solver (Fig. \ref{fig:g3p1}), ii. G3P-2, a cost-effective solver (Fig. \ref{fig:g3p2}), albeit with slower convergence. It is interesting to observe that, among the numerous possibilities allowing for arbitrary cycling from the finest to the coarsest and back to the finest level, the optimization has converged to a V-cycle-like structure. Both of these selected solvers are evaluated across various problem variants and sizes, and their performance is compared to that of standard \ac{AMG} methods.
We assess the performance of our optimized solvers across different values of right-hand sides $f$\footnote{$f$=$rand$ is a vector with random coefficients and unit 2-norm.} and anisotropy in x-direction $a$ for both the original problem size and a larger problem instance (Table \ref{tab:comparewithrefsolvers}). For the original problem size, the optimized solvers G3P-1 and G3P-2 exhibit the fastest and second-fastest solving times, respectively. They outperform reference methods across all problem variants. Notably, G3P-1 demonstrates superior convergence compared to all other solvers. In the case of the larger problem instance, G3P-2 achieves the fastest solving time, while G3P-1 outperforms the reference methods for problems with $f=0$. Despite being the fastest solver for the original problem, G3P-1 is slower than G3P-2 for the larger one. It undergoes a greater deterioration in convergence and being an expensive solver (G3P-1 is picked from the top part of the Pareto front), this affects the total solve time. On the other hand, G3P-2 being more computationally efficient (lying below G3P-1 in the Pareto front) can handle additional iterations with minimal overhead. This highlights the advantages and motivations behind having a pool of Pareto optimal solutions to choose from. Depending on the set of available problems, one could opt for a fast solver either based on its algorithmic efficiency or computational efficiency. 
\begin{table}
\centering
\setlength{\tabcolsep}{2pt}
\begin{tabular}{c||c|c|c|c|c|}
\multirow{2}{*}{} & \multicolumn{5}{c|}{\textbf{(Solve Time (s), Number of Iterations) on a $100\times100\times100$ grid}}\\
\cline{2-6}
 & $f$=$0$$,a$=$0.01$ & $f$=$0$$,a$=$0.001$ & $f$=$0$$,a$=$0.0001$ & $f$=$1$$,a$=$0.001$ & $f$=$rand$$,a$=$0.001$ \\
\cline{2-6}
$V(2,1)$ & $(0.33,10)$ & $(0.33,10)$ & $(0.31,10)$ & $(0.32,10)$ & $(0.26,8)$ \\
$V(3,2)$ & $(0.36,9)$ & $(0.31,8)$ & $(0.3,8)$ & $(0.31,8)$ & $(0.25,6)$ \\
$V(3,3)$ & $(0.35,8)$ & $(0.34,8)$ & $(0.3,7)$ & $(0.31,7)$ & $(0.31,6)$ \\
$G3P$-$1$ & $(0.27,7)$ & $(0.22,6)$ & $(0.22,6)$ & $(0.26,7)$ & $(0.19,5)$ \\
$G3P$-$2$ & $(0.29,10)$ & $(0.28,10)$ & $(0.29,10)$ & $(0.29,10)$ & $(0.24,8)$ \\
\cline{2-6}
\hline
\multirow{2}{*}{} & \multicolumn{5}{c|}{\textbf{(Solve Time (s), Number of Iterations) on a $400\times400\times400$ grid}}\\
\cline{2-6}
 & $f$=$0$$,a$=$0.01$ & $f$=$0$$,a$=$0.001$ & $f$=$0$$,a$=$0.0001$ & $f$=$1$$,a$=$0.001$ & $f$=$rand$$,a$=$0.001$ \\
\cline{2-6}
$V(2,1)$ & $(31.21,16)$ & $(22.27,12)$ & $(18.89,10)$ & $(27.27,13)$ & $(15.24,8)$ \\
$V(3,2)$ & $(36.7,14)$ & $(23.95,10)$ & $(19.96,8)$ & $(26.18,11)$ & $(18.28,7)$ \\
$V(3,3)$ & $(36.35,13)$ & $(29.19,10)$ & $(21.44,8)$ & $(26.76,10)$ & $(16.83,6)$ \\
$G3P$-$1$ & $(27.3,11)$ & $(20.73,8)$ & $(16.88,7)$ & $(30.36,12)$ & $(20.91,8)$ \\
$G3P$-$2$ & $(19.07,11)$ & $(18.15,11)$ & $(17.54,11)$ & $(20.24,11)$ & $(13.27,8)$ \\
\cline{2-6}
\hline
\end{tabular}
\caption{Comparing the solve time and number of iterations of optimized solvers with standard reference methods for different right-hand sides and anisotropy values on the original problem size (top) and a larger instance (bottom), both evaluated on a
28-core Intel Xeon CLX-8276L processor with 28 OpenMP threads.}
\label{tab:comparewithrefsolvers}
\end{table}
\\ Next, we perform a weak scaling study to assess the performance of the solvers on different problem sizes such that the global number of unknowns $N=np \times np \times np$, where $P=p^3$ is the total number of processes. $f$=$0,a$=$0.001$ and the local problem size  $n^3=20^3$. $P$ ranges from 216 to 2744 processes distributed across 49 nodes of the Ruby compute cluster using MPI (1 rank per core). We compare the performance of the optimized solvers with $V(2,1)$ cycles, which scaled the best among the reference methods. $V(2,1)$ employs two sweeps of Gauss-Seidel forward solve on the down cycle, and one sweep of Gauss-Seidel backward solve on the up cycle. Notably, both G3P-1 and G3P-2 are almost twice as fast compared to a $V(2,1)$ solver and also converge with a lesser number of iterations (Fig. \ref{fig:weakscaling_poisson}). Also, inspite of being optimized for a fixed problem size, both G3P methods scale well algorithmically (in iteration count). 
\subsection{AMG as a preconditioner}
\begin{figure} 
  \centering
  \includegraphics[width=\textwidth]{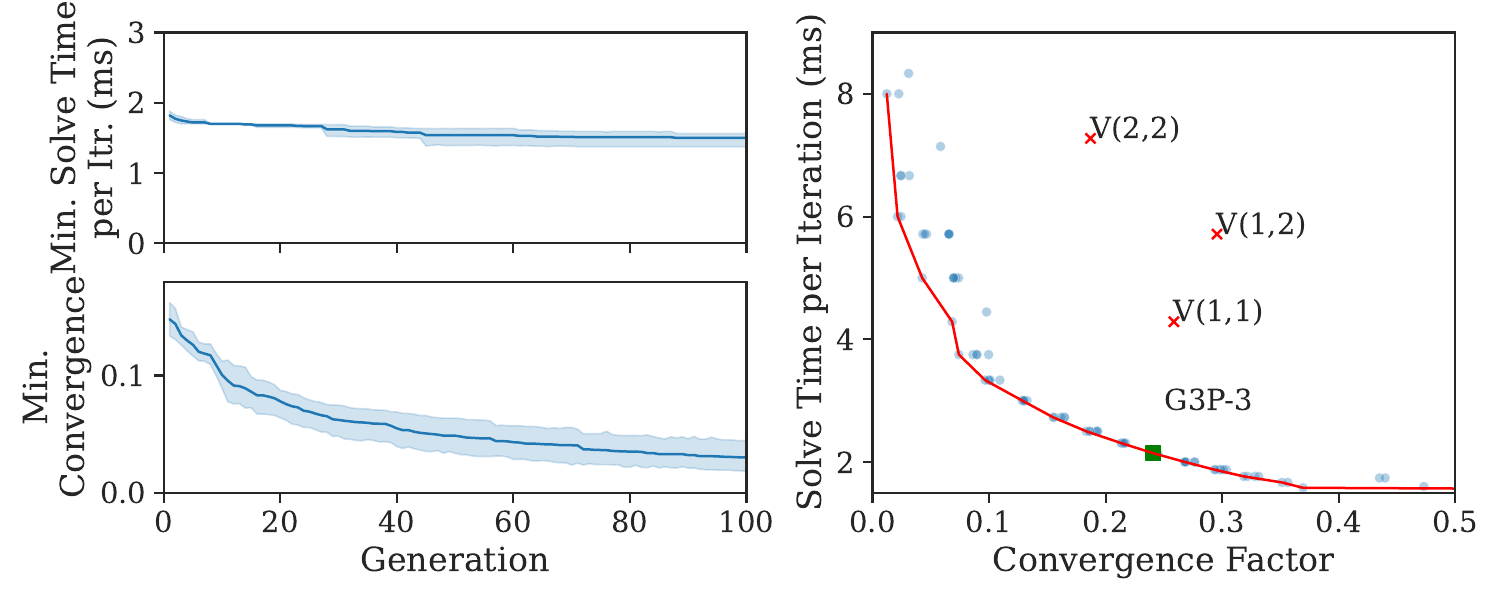}
  \caption{Mean and standard deviation of the minimum objective function values during the optimization (left) and combined Pareto distributions (right), over five experiments.
}
  \label{fig:amgprecond_optimization}
\end{figure}

\begin{figure}[H]
    \centering
    \begin{subfigure}{1\textwidth}
        \centering
        \includegraphics[width=0.9\linewidth]{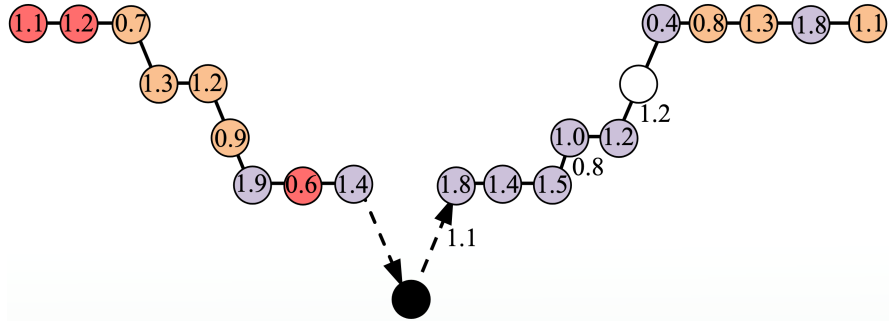}
        \caption{G3P-1}
        \label{fig:g3p1}
    \end{subfigure}
    
    \vspace{1em} 
    
    \begin{subfigure}{0.45\textwidth}
        \centering
        \includegraphics[width=\linewidth]{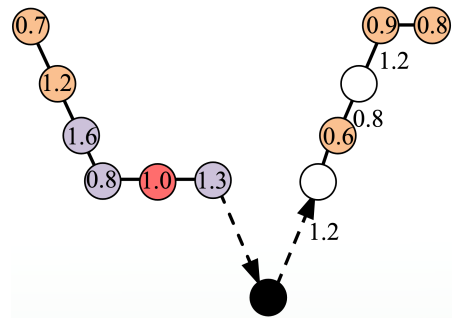}
        \caption{G3P-2}
        \label{fig:g3p2}
    \end{subfigure}
    
    \vspace{1em} 
    
    \begin{subfigure}{0.7\textwidth}
        \centering
        \includegraphics[width=\linewidth]{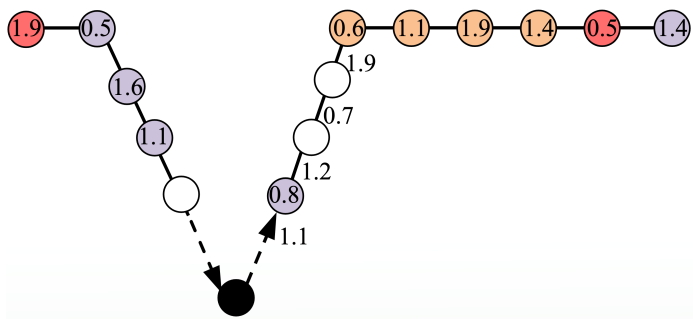}
        \caption{G3P-3}
        \label{fig:g3p3}
    \end{subfigure}
    
    \caption{The structure of the \ac{G3P} methods. The color of the node denotes the type of operation. Yellow: Gauss-Seidel forward solve, Red: Gauss-Seidel backward solve, Purple: Jacobi, Black: Gauss Elimination, White: No Operation. The relaxation factors are indicated inside the node for smoothing and on the edges for coarse-grid correction. The dotted arrow lines represent a V-cycle until the coarsest level.}
    \label{fig:gp3solvers}
\end{figure}

Here, we optimize \ac{AMG} preconditioners for a \ac{CG} method. An \ac{AMG-PCG} method is used for solving a time-stepping multiphysics simulation from the Ares framework. The equation governing the simulation is given by $\frac{\partial\phi}{\partial t} + \nabla \cdot \vec{F} = \mathcal{S}$, where $\vec{F}$ adheres to Fick's Law, expressed as $\vec{F} = -D\nabla\phi$, and $\mathcal{S}$ represents the source term(s). The spatial domain is a 3D block featuring reflecting boundary conditions on the four faces ($xz, xy$) and open boundary conditions on the remaining $yz$ faces. The system is stimulated by a 1 keV temperature source applied at the bottom $yz$ face at $t=0$. The problem has 64000 unknowns and the \ac{AMG} method operates on a grid hierarchy of 7 levels. One cycle of \ac{AMG} is used per \ac{CG} iteration and the system is considered solved when the relative residual norm is less than $10^{-8}$. \\ We consider the system derived from time step $t=1$ for the optimization. The objectives in the optimization are now measured for the PCG solver using different \ac{AMG} preconditioners. We observe that the minimum value of the objectives in the population decreases during the optimization (Fig. \ref{fig:amgprecond_optimization}, left). But the total time to solution for \ac{AMG-PCG} is minimized when both objectives are close to optimum concurrently. Upon concluding the optimization after 100 generations, we obtain a collection of Pareto-optimal \ac{AMG-PCG} methods (Fig. \ref{fig:amgprecond_optimization}, right). \ac{AMG} preconditioners employing standard \ac{AMG} cycles \footnote{Standard cycles use Gauss-Seidel forward solve for pre-smoothing, and Gauss-Seidel backward solver for post-smoothing. Among the choice of smoothers considered for the optimization, this was found to be the most optimal.} are positioned outside the Pareto front, indicating their sub-optimal nature. We choose a well-performing \ac{AMG} preconditioner, G3P-3 (Fig. \ref{fig:g3p3}) from the Pareto front.
The optimized \ac{AMG-PCG} method is applied to solve systems obtained from different time instances within the same simulation (Table \ref{tab:comparewithrefsolvers_ares}). The aim is to assess the robustness of the optimized solver when dealing with variations in the system matrices, as these matrices change during the simulation. We observe that G3P-3, optimized for $t=1$, performs effectively across a range of time steps $t=2,3,4...,10$. G3P-3 preconditioned \ac{CG} outperforms PCG with standard \ac{AMG} preconditioners for all time steps, except for $t=2,5,7,8$ wherein a $V(1,1)$ \ac{AMG} preconditioner is equally effective. 
\begin{table}
\centering
\begin{tabular}{c||*{4}{p{1.5cm}|}}
\multirow{2}{*}{} & \multicolumn{4}{c|}{\textbf{(Solve Time (ms), Number of \ac{CG} Iterations)}}\\
\cline{2-5}
 $t$ & $V(1,1)$ & $V(1,2)$& $V(2,2)$ & $G3P$-$3$ \\
\cline{2-5}
 $1$ & $(60, 14)$ & $(90, 14)$ & $(70, 11)$ & $(50, 14)$ \\
        $2$ & $(50, 11)$ & $(60, 10)$ & $(70, 9)$  & $(50, 12)$ \\
        $3$ & $(60, 12)$ & $(70, 11)$ & $(70, 9)$  & $(50, 13)$ \\
        $4$ & $(50, 11)$ & $(60, 9)$  & $(60, 8)$  & $(40, 12)$ \\
        $5$ & $(50, 11)$ & $(70, 11)$ & $(70, 9)$  & $(50, 13)$ \\
        $6$ & $(50, 9)$  & $(50, 8)$  & $(60, 7)$  & $(40, 8)$  \\
        $7$ & $(50, 11)$ & $(60, 11)$ & $(60, 9)$  & $(50, 12)$ \\
        $8$ & $(40, 9)$  & $(50, 8)$  & $(60, 7)$  & $(40, 9)$  \\
        $9$ & $(60, 12)$ & $(60, 11)$ & $(70, 9)$  & $(40, 13)$ \\
        $10$ & $(50, 8)$  & $(60, 8)$  & $(50, 6)$  & $(30, 8)$  \\
\cline{2-5}
\hline
\end{tabular}
\caption{Comparing the solve time and number of \ac{CG} iterations of optimized preconditioners with standard \ac{AMG} cycles at different time instances,  evaluated on a 28-core Intel Xeon CLX-8276L processor with 28 OpenMP threads.}
\label{tab:comparewithrefsolvers_ares}
\end{table}

\FloatBarrier
        
\section{Conclusion}\label{sec:conc}
 We used a novel grammar-guided approach for the automated generation of \ac{AMG} methods, utilizing a set of grammar rules. These generated methods are not confined to standard cycle structures but instead possess arbitrary cycling capabilities. This flexibility was harnessed to search for more efficient \ac{AMG} methods using \ac{G3P}. Our results indicate that \ac{AMG} methods with flexible cycles specifically tailored for a given problem are superior in performance when compared to \ac{AMG} methods with standard cycles. They are also seen to generalize well for different problem variants and sizes when used as a standalone solver. Furthermore, when optimized as a preconditioner in a time-stepping simulation code, they maintain their efficiency across different time instances. However, we have not delved into an in-depth analysis of the optimized solvers to gain insights into why the evolved cycle structures prove to be efficient. Also, it is not yet fully understood which components (smoothers, relaxation weights, cycle structures), exert the most influence on optimality. Another potential future direction could be formulating additional grammar rules that include the \ac{AMG} setup phase within the optimization process. Furthermore, the collection of Pareto optimal solutions generated could serve as a valuable repository with a rich dataset of diverse AMG methods. One possible use of this dataset could be to train a machine learning model, which, in turn, could be used to select \ac{AMG} methods tailored to specific problem instances.
\section{Acknowledgements}
This work was performed under the auspices of the U.S. Department of Energy by Lawrence Livermore National Laboratory under Contract DE-AC52-07NA27344. Additionally, this work has received funding from the European High Performance Computing Joint Undertaking (JU) and Sweden, Germany, Spain, Greece, and Denmark under grant agreement No 101093393.

\newpage
\bibliographystyle{siamplain}
\bibliography{references}
\end{document}